\documentclass[a4paper,11pt]{article}
\usepackage{longtable}
\usepackage{xltabular}
\pdfoutput=1 

\usepackage{jcappub} 

\usepackage[T1]{fontenc} 

\usepackage{xcolor}

\title{\boldmath Cosmological-model independent limits on photon mass from FRB and SNe data}

\author[a,1]{Thais Lemos,\note{Corresponding author}}
\author[b,a]{Rodrigo Gon\c{c}alves,}
\author[a]{Joel Carvalho,}
\author[a]{Jailson Alcaniz}

\affiliation[a]{Observat\'orio Nacional, Rio de Janeiro - RJ, 20921-400, Brasil}
\affiliation[b]{ Departamento de F\'{\i}sica, Universidade Federal Rural do Rio de Janeiro, Serop\'edica - RJ, 23897-000, Brasil}

\emailAdd{thaislemos@on.br}
\emailAdd{rsg\_goncalves@ufrrj.br}
\emailAdd{jcarvalho@on.br}
\emailAdd{alcaniz@on.br}

\abstract{Electromagnetic emissions from astrophysical sources at cosmological distances can be used to estimate the photon mass, $m_{\gamma}$. In this paper, we combine measurements of the dispersion measure ($\mathrm{DM}$) of fast radio bursts (FRB) with the luminosity distance from type Ia supernovae (SNe) to investigate update constraints on the photon rest mass. We derive the expression of  $\mathrm{DM}$ dependence concerning a non-vanishing photon mass from a cosmological-model independent approach and constrain the parameter $m_{\gamma}$ from measurements of 68 well-localized FRBs and 1048 SNe data from the Pantheon compilation. We consider two scenarios for the baryon fraction in the intergalactic medium ($f_{\mathrm{IGM}}$): one where the value is fixed according to recent reports and another where it is treated as a free parameter, $f_{\mathrm{IGM}} = f_{\mathrm{IGM,0}}$. In the latter case, we find $m_{\gamma} = (29.4_{-15.5}^{+5.80}) \times 10^{-51}$ kg, at $1\sigma$ level. Our results also demonstrate an anticorrelation between $f_{\mathrm{IGM}}$ and $m_{\gamma}$, which highlights the importance of analyzing a larger sample of FRBs for a more comprehensive understanding of their properties.}

\begin{document}
\maketitle
\flushbottom

\section{Introduction}
\indent

The invariance of the speed of light is a fundamental postulate of  Einstein’s special relativity \cite{Einstein1905}. This postulate implies that photons should be massless so that evidence for a nonzero photon mass would imply new physics theories beyond relativity, such as the de Broglie-Proca theory \cite{Broglie1922,Proca1936} or models of massive photons as an explanation for dark energy \cite{Kouwn2016}. Within that context, constraining the photon mass, $m_\gamma$, is an effective way to test the validity of this postulate. With this purpose, several experiments and observations have been developed in order to put tight constraints on photon mass, such as Coulomb's inverse square law \cite{Williams1971}, the Cavendish torsion balance \cite{Lakes1998,Luo2003}, gravitational deflection of electromagnetic waves \cite{Lowenthal1973}, Jupiter’s magnetic field \cite{Davis1975}, pulsar spindown \cite{Yang2017}, and so on. In contrast, the test based on frequency-dependent dispersion of light is more robust \cite{Tu2004,Tu2005}.

The massive photon group velocity depends on the photon energy, meaning that higher-energy photons travel faster than lower-energy ones. Thus, if two photons with different frequencies are emitted simultaneously from the same source, they will reach Earth at different times, leading to measurable time delays in signals emitted from astrophysical sources (see reference \cite{Wei2021} for a review). In this context, Fast Radio Bursts (FRBs) provide one of the ideal laboratories for testing aspects of fundamental physics due to observed time delays between correlated particles or photons. 

FRBs are a class of millisecond transient events observed in radio frequency ($\sim$ GHz) (see \cite{Thornton2013,Petroff2015,Petroff2016,Platts2019,Petroff2022} for a review of FRBs). The first burst was discovered in 2007 by the Parkes Telescope \cite{Lorimer2007} and since then more than a thousand events have been detected by new surveys, such as e.g. the Canadian Hydrogen Intensity Mapping Experiment (CHIME, \cite{CHIME}). In recent years, several models have been proposed to explain the radiation mechanism \cite{Zhang2023} and the progenitor source \cite{Platts2019} of these events, but both remain unknown. The large observed dispersion measure (DM) with respect to the Milky Way contribution indicates an extragalactic or cosmological origin for the FRB. The total DM of FRBs is the integral of the column density of free electrons along the line of sight and makes it possible to use extragalactic FRBs for cosmological investigations. Although there are only a few FRBs with the redshift of the host galaxy (well-localized) in the literature, one uses it as an astrophysical and cosmological probe combining DM with the redshift (see references \cite{Walters2018,Wei2018,Lin2021,Wu2021,Reischke2023,Lemos2023,Lemos2025} for cosmological applications). 

In practice, some issues hinder the application of FRBs for cosmological purposes. For instance, the variance in the dispersion measure relates to the inhomogeneous cosmic distribution of the electrons \cite{Takahashi2021}. These density fluctuations ($\delta$) can be treated as a probability distribution \cite{Macquart2020} or as a fixed value in the statistical analysis \cite{Takahashi2021}. Another limitation in the analysis is the contribution of the host galaxy ($\mathrm{DM}_{\mathrm{host}}$) because it is a complex parameter to observe and measure \cite{Xu2015}. To circumvent this problem, $\mathrm{DM}_{\mathrm{host}}$ can be assumed to be a free parameter or a log-normal distribution \cite{Walker2020}. Finally, the last restriction is the poor knowledge about the evolution of the fraction of baryon mass in the intergalactic medium ($f_{\mathrm{IGM}}$) \cite{Shull2012,Meiksin2009}, which is degenerated with the cosmological parameters.

Some works in the literature have constrained the photon mass from FRBs within the framework of a flat $\Lambda$CDM cosmological model \cite{Wu2016,Bonetti2016,Bonetti2017,Shao2017,Xing2019,Wei2020,Wang2021,Chang2023,Lin2023,Wang2023}. However, it is important to note that the $\Lambda$CDM model is built on General Relativity (GR), in which photons are massless particles. Consequently, employing this model introduces a circularity issue. In contrast, the authors in Reference \cite{Ran2024} adopt a model-independent approach to constrain the photon's rest mass, combining FRBs with Hubble parameter measurements as well as fixing the $f_{\mathrm{IGM}}$ at a given value, and they found $m_{\gamma} \leq 3.5 \times 10^{-51}$ kg (1$\sigma$). 

In a previous work \cite{Lemos2023}, we proposed a cosmological model-independent method to constrain the baryon fraction in the intergalactic medium ($f_{\mathrm{IGM}}$) from FRBs observations. Here, we apply this method to constrain $m_{\gamma}$. We combine a sample of 68 well-localized FRBs with type Ia supernovae observations taking into account a significant anti-correlation between $f_{\mathrm{IGM}}$ and $m_{\gamma}$. This paper is organized as follows. Section \ref{sec:DM} summarizes the concepts of time delay and dispersion measures of FRBs. We present our method to derive the dispersion measure components in Section \ref{sec:photon_framework}, while Section \ref{sec:methodology} presents the data set used and our methods. In Section \ref{sec:results} we present the results of our analysis and discuss the role of $f_{\mathrm{IGM}}$ in the results by setting it as a fixed and free parameter. We end the paper by summarizing the main conclusions in Section \ref{sec:conclusions}.

\section{Dispersion of FRBs}\label{sec:DM}
\indent

The radio waves from the FRB source to the Earth can suffer dispersion in the plasma, causing the high-frequency photons to arrive earlier than the low-energy ones. So in this section, we will summarize the well-established concepts of the FRBs’ time delay and dispersion measure (see Ref. \cite{Rybicki}). 

\subsection{Definition of Dispersion Measure}
\indent

The time required for a wave from a source at a distance $d$ with frequency $\omega$ and group velocity $v_{g}$ to reach Earth can be written as \cite{Rybicki}

\begin{equation}
    t_{p} = \int_{0}^{d} \frac{dS}{v_{g}} \approx \int_{0}^{d} c^{-1} \left( 1 + \frac{\omega^{2}_{p}}{2\omega^{2}} \right) dS\;,
\end{equation}
where $S$ measures the line-of-sight distance from the source to the Earth, and the plasma frequency is given by 

\begin{equation}\label{eq:wp}
    \omega_{p}^{2} = \frac{4\pi n_{e}e^{2}}{m_{e}}\;,
\end{equation}
being $n_{e}$, $e$, $m_{e}$ the density of free electrons, the electron charge, and the electron mass, respectively. If the frequency does not change along with $S$, then

\begin{equation}\label{eq:t_p}
    t_{p} \approx \frac{d}{c} + \left(2\omega^{2}c \right)^{-1}\int_{0}^{d}\omega^{2}_{p} dS\;. 
\end{equation}
The quantity that is usually measured is the rate change of the arrival time with respect to frequency. Replacing the plasma frequency relation (Eq. \ref{eq:wp}) into Eq. \ref{eq:t_p} and assuming that the density of free electrons is the only quantity that changes with $S$, we find

\begin{equation}
    \frac{dt_{p}}{d\omega} = - \frac{4\pi e^{2}}{\omega^{3}m_{e}c} \int_{0}^{d} n_{e} \; dS\;,
\end{equation}
where 

\begin{equation}\label{eq:dm}
    \mathrm{DM} = \int_{0}^{d} n_{e} \; dS\;
\end{equation}
is the dispersion measure.

\subsection{Cosmological Extension}
\indent

Assuming that the FRB comes from an extragalactic source at redshift $z$, it is necessary to take into account three effects \cite{Zheng_2014}: the change in the observed frequency from the redshift of light ($\omega \rightarrow \omega = (1+z)\omega_{\mathrm{obs}}$); a redshift dependence on the electronic density ($n_e = n_e(S) \rightarrow n_e = n_e(z)$); and the time dilation effect (with the introduction of a $(1+z)$ term). The distance of the propagated electromagnetic wave is $dS = cdt$ while the time variation can be obtained from the definition of the Hubble parameter $dt = - \frac{dz}{(1+z)H(z)}$. Thus, the rate change of the arrival time (Eq. \ref{eq:t_p}) becomes


\begin{equation}\label{eq:tp_cosmo}
t_{p} = \frac{d}{c} + \frac{1}{2\omega_{\mathrm{obs}}^{2}} \int_{0}^{z} \frac{\omega^{2}_{p}}{(1+z)^2H(z')}dz'\;,
\end{equation}
whose derivative with respect to the observed frequency is given by

\begin{equation}
\label{eq:dt/dw}
    \frac{dt_{p}}{d\omega_{\mathrm{obs}}} = - \frac{1}{\omega^{3}_{\mathrm{obs}}} \int_{0}^{z}  \frac{\omega^{2}_{p}}{(1+z)^2H(z')}dz'\;.
\end{equation}
Now, replacing the plasma frequency (Eq. \ref{eq:wp}), we obtain

\begin{equation}\label{eq:dt/dwSt}
    \frac{dt_{p}}{d\omega_{\mathrm{obs}}} = - \frac{4\pi e^{2}}{\omega^{3}_{\mathrm{obs}}m_{e}c} \int_{0}^{z} \frac{n_e(z')c}{(1+z')^{2}H(z')} dz'\;,
\end{equation}
where the speed of light in the numerator and denominator is necessary to guarantee the correct units of the dispersion measure, which now is written as

\begin{equation}\label{eq:DM_cosmo}
    \mathrm{DM} = \int_{0}^{z}   \frac{n_e(z')c}{(1+z')^{2} H(z')}dz'\;.
\end{equation}

\section{Non vanishing photon mass framework}\label{sec:photon_framework}
\indent

\subsection{Time delay}

Let us consider a scenario where the photons have a small but nonzero rest mass. In this case, the dispersion relation considering both nonzero photon mass ($m_{\gamma}$) and effects of plasma dispersion is different from the usual and can be written as \cite{Proca1936,Goldhaber} 

\begin{equation}
    \omega^{2} = k^{2}c^{2} + \omega_{p}^{2} + \omega_{\gamma}^{2} \;,
\end{equation}
where the frequency related to the photon mass is

\begin{equation}\label{eq:w_gamma}
    \omega_{\gamma}^{2} = \frac{m_{\gamma}^{2}c^{4}}{\hbar^{2}} \;.
\end{equation}
The group velocity of the electromagnetic wave is \cite{Goldhaber} 

\begin{equation}\label{eq:vg}
    v_{g} = \frac{d\omega}{dk} = c  \Bigg( 1 - \frac{\omega_{p}^{2} + \omega_{\gamma}^{2} }{2\omega^{2}}\Bigg) \;. 
\end{equation}
while the time required for a radio wave to reach the Earth from a source at a distance $d$ and group velocity given by Eq. \ref{eq:vg} is

\begin{equation}\label{eq:t_p_gamma}
    t_{p} \approx \int_{0}^{d} c^{-1} \Bigg( 1 + \frac{\omega_{p}^{2} + \omega_{\gamma}^{2} }{2\omega^{2}}\Bigg) dS \;.
\end{equation}
Note that the only difference between the expression above and Eq. \ref{eq:t_p} is the term related to photon mass. So, it is possible to write the total time delay as the contribution from plasma dispersion and non-vanishing photon mass

\begin{equation}\label{eq:time_delay}
    \Delta t_{\mathrm{obs}} = \Delta t_{\mathrm{DM}} + \Delta t_{\gamma} \;.
\end{equation}

As done before, we will consider that the FRB signal comes from a source at redshift $z$. In this case, Equation \ref{eq:t_p_gamma} becomes 

\begin{equation}
    t_{p} =  \int_{0}^{z} \Bigg( 1 + \frac{\omega_{p}^{2} + \omega_{\gamma}^{2} }{2\omega_{\mathrm{obs}}^{2} (1+z')^{2}}\Bigg) \frac{dz'}{H(z')} \;.
\end{equation}
Replacing Eqs. \ref{eq:wp} and \ref{eq:w_gamma} into the expression above and calculating the rate change of the arrival time, we obtain

\begin{equation}
\frac{dt_{p}}{d\omega_{\mathrm{obs}}} = - \frac{4\pi e^{2}}{m_{e}\omega_{\mathrm{obs}}^{3}} \int_{0}^{z} \frac{n_{e}(z')}{(1+z')^{2}} \frac{dz'}{H(z')} - \frac{1}{\omega_{\mathrm{obs}}^{3}} \int_{0}^{z} \frac{m_{\gamma}^{2}c^{4}}{\hbar^{2}(1+z')^{2}} \frac{dz'}{H(z')} \;,
\end{equation}
or still

\begin{equation}\label{eq:dt_dw}
\frac{dt_{p}}{d\omega_{\mathrm{obs}}} = - \frac{4\pi e^{2}}{m_{e}\omega_{\mathrm{obs}}^{3}c} \int_{0}^{z} \frac{n_{e}(z')c}{(1+z')^{2}} \frac{dz'}{H(z')} - \frac{1}{\omega_{\mathrm{obs}}^{3}} \frac{e^{2}}{\varepsilon_{o}m_{e}c} \int_{0}^{z}  \frac{\varepsilon_{o}m_{e}c}{e^{2}}\frac{m_{\gamma}^{2}c^{4}}{\hbar^{2}(1+z')^{2}} \frac{dz'}{H(z')} \;,\end{equation}
where $\varepsilon_{o}$ is the permittivity of vacuum. Note that the first integral in Eq. \ref{eq:dt_dw} is the dispersion measure relation (Eq. \ref{eq:DM_cosmo}) and the second term we can be defined as the equivalent of DM arising from the massive photons 

\begin{equation}\label{eq:DMgamma}
    \mathrm{DM}_{\gamma} (z) = \frac{\varepsilon_{o}m_{e}c^{5}}{\hbar^{2}e^{2}}m_{\gamma}^{2} \int_{0}^{z}  \frac{(1+z')^{-2}}{H(z')} dz' \;.
\end{equation}

\subsection{Dispersion Measure Components}
\indent 

In what follows, we will discuss the components of the dispersion measure considering the nonvanishing photon rest mass scenario. Radio waves can be dispersed in the plasma due to the interaction of photons from the source with the free electrons in the medium through which the wave passes. For this reason, the received radio waves experience a change in frequency, which in turn induces a change in their speed. Therefore, we can write the observed dispersion measure ($\mathrm{DM}_\mathrm{obs}$) as a contribution of several components \cite{Deng2014,Gao2014} including the term related to massive photons

\begin{equation}
\label{eq:DMobs}
\mathrm{DM}_{\mathrm{obs}}(z) = \mathrm{DM}_{\mathrm{ISM}} + \mathrm{DM}_{\mathrm{halo}} + \mathrm{DM}_{\mathrm{host}}(z) + \mathrm{DM}_{\mathrm{IGM}}(z) + \mathrm{DM}_{\gamma}(z) \;,
\end{equation}
where the subscripts ISM, halo, host, and IGM denote contributions from the Milky Way Interstellar Medium, the Milky Way Halo, the FRB host galaxy, and the Intergalactic Medium, respectively. The quantity $\mathrm{DM}_{\mathrm{obs}}(z)$ of a FRB is directly measured from the corresponding event. The component $\mathrm{DM}_{\mathrm{ISM}}$ can be well-constrained using models of the ISM galactic electron distribution in the Milky Way from pulsar observations \cite{Taylor1993,Cordes2002,Yao2017}, while the Milky Way halo contribution is not well-constrained yet. In the present paper, we follow \cite{Macquart2020} and assume $\mathrm{DM}_{\mathrm{halo}} = 50$ pc/cm$^{3}$ . 

To perform a statistical comparison between the observational data and the theoretical parameters, we define the observed extragalactic dispersion measure as

\begin{equation}
\label{eq:DMext_obs}
  \mathrm{DM}^{\mathrm{obs}}_{\mathrm{ext}}(z) \equiv  \mathrm{DM}_{\mathrm{obs}}(z) - \mathrm{DM}_{\mathrm{ISM}}  - \mathrm{DM}_{\mathrm{halo}} \; ,
\end{equation}
and the theoretical extragalactic dispersion measure as
\begin{equation}
\label{eq:DMext_th}
    \mathrm{DM}_{\mathrm{ext}}^{\mathrm{th}}(z) \equiv \mathrm{DM}_{\mathrm{host}}(z) + \mathrm{DM}_{\mathrm{IGM}}(z) + \mathrm{DM}_{\gamma}(z) \; ,
\end{equation}
where one can note that $\mathrm{DM}_{\mathrm{host}}(z)$, $\mathrm{DM}_{\mathrm{IGM}}(z)$ and $\mathrm{DM}_{\gamma}(z)$ depend on $\mathrm{DM}_{\mathrm{host,0}}$, $f_{\mathrm{IGM,0}}$ and $m_{\gamma}$, respectively. Thus, changes in $m_{\gamma}$ will lead to changes in the other free parameters, $f_{\mathrm{IGM,0}}$ and $\mathrm{DM}_{\mathrm{host,0}}$. 

The host galaxy component is not well-constrained due to the challenges in modeling and measuring it. These difficulties happen because that component depends on many features, such as the type of galaxy, the relative orientations of the FRBs source concerning the host and source, and the near-source plasma \cite{Xu2015}. For this reason, we can write the redshift evolution of $\mathrm{DM}_{\mathrm{host}}(z)$ using the relation  \cite{Deng2014, Ioka2003}

\begin{equation}
\label{DMhost}
\mathrm{DM}_{\mathrm{host}}(z) = \frac{\mathrm{DM}_{\mathrm{host},0}}{(1+z)}\;,
\end{equation}
where the $(1+z)$ factor accounts for the cosmological time dilation for a source at
redshift $z$.

The largest component of $\mathrm{DM}_{\mathrm{obs}}$ is the average dispersion measure from the IGM, where cosmological contributions appear. Replacing the electronic density $n(z)$, as given by \cite{Deng2014}, in the first integral of Eq. \ref{eq:dt_dw}, we obtain 

\begin{equation}\label{eq:DMigm}
\mathrm{DM}_{\mathrm{IGM}}(z) = \frac{3c\Omega_{b}H_{0}^{2}}{8\pi Gm_{p}} \int_{0}^{z} \frac{(1+z')f_{\mathrm{IGM}}(z')\chi(z')}{H(z')}  dz'\;,
\end{equation}
where $\Omega_{b}$ is the present-day baryon density parameter, $H_{0}$ is the Hubble constant, $m_{\gamma}$ is the proton mass, $f_{\mathrm{IGM}}(z)$ is the baryon fraction in the IGM, and $\chi(z)$ is the free electron number fraction per baryon. This last quantity is given by
$\chi(z) = Y_{H} \chi_{e,H}(z) + Y_{He} \chi_{e,He}(z)$, 
where the terms $Y_{H} = 3/4$ and $Y_{He} = 1/4$ are the mass fractions of hydrogen and helium, respectively, while $\chi_{e,H}(z)$ and $\chi_{e,He}(z)$ are the ionization fractions of hydrogen and helium, respectively. At $z < 3$ hydrogen and helium are fully ionized ($\chi_{e,H}(z) = \chi_{e,He}(z) = 1$) \cite{Becker2011}, so that we have $\chi(z) = 7/8$.

\section{Data and Methodology}\label{sec:methodology}
\indent

\subsection{Data: FRBs}
\indent 

The currently available sample of FRBs contains 97 well-localized events (for details of FRBs catalog\footnote{https://blinkverse.alkaidos.cn}, see \cite{Blinkverse}). For the reasons below, we excluded from our analysis the following events: FRB 20171020A \cite{FRB171020A} and FRB 20181030 \cite{Bhardwaj2021_2} are at $z = 0.0087$ and $z = 0.0039$, respectively, and can not be associated with any SNe Ia in the Pantheon catalog; FRB 20190520B \cite{FRB190520B} has a host galaxy contribution significantly larger than the other events; FRB 20190614 \cite{FRB190614} has no measurement of spectroscopic redshift and can be associated with two host galaxies; FRB 20200120E \cite{Bhardwaj2021} is estimated close to M81 at a distance of $\sim 3.6$ Mpc, but a Milky Way halo origin can not be rejected; FRB 20210405I \cite{FRB210405I} and FRB 20220319D \cite{FRB220207C} have the MW contribution larger than observed one ($\mathrm{DM}_{\mathrm{ext}}^{\mathrm{obs}} < 0$); and finally, the events FRB 20220529A \cite{FRB220529A}, FRB 20221027A, FRB 20221029A, FRB 20221101B, FRB 20221113A, FRB 20221116A, FRB 20230124, FRB 20230216A, FRB 20230307A, FRB 20230501A, FRB 20230521B, FRB 20230626A, FRB 20230628A, FRB 20230712A, FRB 20231120A, FRB 20231123B, FRB 20231220A, FRB 20240119A, FRB 20240123A, FRB 20240213A, FRB 20240215A, FRB 20240229A \cite{FRB220204A} have no uncertainty associated to the $\mathrm{DM}_{\mathrm{obs}}$ available. In Table \ref{tab:data} we list our working sample that contains 68 FRBs \cite{FRB121102,FRB20191228,FRB180814,FRB180916,FRB180924,FRB181112,FRB181220A,FRB190102,FRB190110C,FRB190523_1,FRB190523_2,FRB190608,FRB201123A,FRB201124,FRB210117A,FRB210320,FRB210410D,FRB210603A,FRB210807D,FRB211203C,FRB220204A,FRB220207C,FRB20220610A,FRB220717A,FRB220912A,FRB221219A,FRB240114A}, including their main properties: redshift, the Galaxy contribution ($\mathrm{DM}_{\mathrm{MW,ISM}}$) estimated from the NE2001 model \cite{Cordes2002}, observed dispersion measure ($\mathrm{DM}_{\mathrm{obs}}$), $\mathrm{DM}_{\mathrm{obs}}$ uncertainty ($\sigma_{\mathrm{obs}}$) and the references. 

The observational quantity $\mathrm{DM}_{\mathrm{ext}}$ can be estimated from Table \ref{tab:data} and its total uncertainty can be expressed by the relation

\begin{equation}\label{eq:uncertainty}
    \sigma_{\mathrm{tot}}^{2} = \sigma_{\mathrm{obs}}^{2} + \sigma_{\mathrm{MW}}^{2} + \sigma_{\mathrm{IGM}}^{2} + \bigg( \frac{\sigma_{\mathrm{host},0}}{1+z} \bigg)^{2} + \delta^{2} \;,
\end{equation}
where $\sigma_{\mathrm{obs}}$ is listed in Table \ref{tab:data}, $\sigma_{\mathrm{MW}}$ is the average galactic uncertainty and is assumed to be $10$ pc/cm$^{3}$ \cite{Manchester2005}. Following \cite{Li2019}, we adopt $\sigma_{\mathrm{host},0} = 30$ pc/cm$^{3}$ as the uncertainty of $\mathrm{DM}_{\mathrm{host},0}$ while the fluctuations of $\mathrm{DM}$ are given by $230\sqrt{z}$ pc/cm$^{3}$ \cite{Takahashi2021,Lemos2023}. Furthermore, the uncertainty of IGM contribution ($\sigma_{\mathrm{IGM}}$) can be calculated from error propagation of Eq. \ref{eq:constant}, i.e., 

\begin{eqnarray}
    \sigma_{\mathrm{IGM}} = A f_{\mathrm{IGM},0} \bigg[ \frac{\sigma_{d_{L}}^{2}}{c^{2}} + \frac{\sigma_{\mathrm{I}}^{2}}{c^{2}} \bigg]^{1/2},
\end{eqnarray}
where $\sigma_{d_{L}}$ is the error of luminosity distance, calculated from SNe observations, and $\sigma_{\mathrm{I}}$ is the uncertainty of the quantity given by Eq. \ref{eq:Sum}. 

\subsection{Data: Type IA Supernovae}
\indent 

We use the Pantheon catalog \cite{Scolnic} for the SNe observations, which comprises 1048 SNe within the redshift range $0.01 < z < 2.3$. From the distance moduli relation, 
\begin{equation} \label{eq:mz}
    \mu(z) = m_{B} - M_{B} = 5\log_{10}\left[ \frac{d_{L}(z)}{1\mbox{Mpc}}\right] + 25 \;,
\end{equation}
we can obtain $d_{L}$, where $m_{B}$ and $M_{B}$ are the apparent and absolute magnitude, respectively. In the present work we fix $M_{B} = -19.214 \pm 0.037$ mag \cite{Riess2019}. From the above equation, we can determine the luminosity distance uncertainty

\begin{equation}
    \sigma_{d_{L}} = \frac{\ln{10}}{5}d_{L} \cdot \sqrt{\sigma_{m_{B}}^{2} + \sigma_{M_{B}}^{2}},
\end{equation}
where $\sigma_{m_{B}}$ and $\sigma_{M_{B}}$ are the apparent and absolute magnitude uncertainties, respectively. To estimate $d_{L}$ and its uncertainty at the same redshift of FRBs, we perform a Gaussian Process (GP) reconstruction of the Pantheon data, using GaPP python library\footnote{For details of GaPP (https://github.com/astrobengaly/GaPP), see \cite{GaPP}.}. 

\subsection{Methodology}
\indent

In Reference \cite{Lemos2023}, we presented a cosmological model-independent method, which solves the $\mathrm{DM}_{\mathrm{IGM}}$ integral by parts, identifying one of the terms as the luminosity distance ($d_{L}$). We considered a time-dependent and a constant parameterizations for $f_{\mathrm{IGM}}$, and found that the evidence is inconclusive. Therefore, we will only consider the constant case in our analysis. In this scenario, the Dispersion Measure from the intergalactic medium (Eq. \ref{eq:DMigm}) can be written as 

\begin{equation}\label{eq:constant}
    \mathrm{DM}_{\mathrm{IGM}}(z) = A f_{\mathrm{IGM,0}} \left[\frac{d_{L}(z)}{c} - \frac{1}{c}\int_{0}^{z} \frac{d_{L}(z')}{(1+z')} dz' \right],
\end{equation}
where $A=\frac{3c\Omega_{b}H_{0}^{2}}{8\pi Gm_{p}}$ and the internal can be numerically solved as \cite{Holanda_2013}

\begin{equation}\label{eq:Sum}
    \int_{0}^{z} \frac{d_{L}(z')}{(1+z')} dz' =  \frac{1}{2}\sum_{i=1}^{N} \left( z_{i+1}-z_{i}\right)\nonumber \times \left[\frac{d_{L}(z_{i+1})}{(1+z_{i+1})} 
+ \frac{d_{L}(z_{i})}{(1+z_{i})}  \right]\;.
\end{equation}

For the last component in Eq. \ref{eq:DMobs}, a cosmological model is typically assumed to solve the integral for $\mathrm{DM}_{\gamma}$ (Eq. \ref{eq:DMgamma}). In our analyses, we follow the approach presented in \cite{Lemos2023} and integrate Eq. \ref{eq:DMgamma} by parts using the definition of luminosity distance $d_{L}$, allowing us to avoid relying on any specific cosmological model and instead we use the SNe luminosity distance. Thus, Eq. \ref{eq:DMgamma} becomes 

\begin{equation}\label{eq:gamma}
    \mathrm{DM}_{\gamma} (z) = B m_{\gamma}^{2} \Bigg[ \frac{d_{L}(z)}{(1+z)^{3}} + \frac{2}{c} \int_{0}^{z} \frac{d_{L}(z')}{(1+z')^{4}} dz'\Bigg] \;,
\end{equation}
where $B = \frac{\varepsilon_{o}m_{e}c^{5}}{\hbar^{2}e^{2}} $ and the above integral can be also numerically solved as \cite{Holanda_2013}

\begin{equation}\label{eq:Sum2}
\int_{0}^{z} \frac{d_{L}(z')}{(1+z')^{4}} dz' =  \frac{1}{2}\sum_{i=1}^{N} \left( z_{i+1}-z_{i}\right)\nonumber \times \left[\frac{d_{L}(z_{i+1})}{(1+z_{i+1})^{4}} 
+ \frac{d_{L}(z_{i})}{(1+z_{i})^{4}}  \right]\;.
\end{equation}

From the above expressions, one can constrain the rest mass of the photon by modeling the terms in $\mathrm{DM}_{\mathrm{ext}}^{\mathrm{th}}$ quantity (Eq. \ref{eq:DMext_th}) and comparing these theoretical predictions with the observed values of $\mathrm{DM}_{\mathrm{ext}}$. By combining well-localized FRBs with SNe dataset we can obtain constraints on $m_{\gamma}$, $f_{\mathrm{IGM,0}}$ and $\mathrm{DM}_{\mathrm{host,0}}$ in a cosmological model-independent way from the observational data sets described above.

As previously mentioned, the $f_{\mathrm{IGM}}$ is not currently well determined from observations and simulations, and it could be correlated to $m_{\gamma}$ from Eq. \ref{eq:DMext_th}. Therefore, our analysis considers two different approaches for $f_{\mathrm{IGM,0}}$ to study its impact on the constraining of photon mass. First, we follow References \cite{Bonetti2016,Bonetti2017,Shao2017,Wei2020,Wang2021,Chang2023,Lin2023,Ran2024} and adopt the fixed value $f_{\mathrm{IGM,0}}=0.83$ \cite{Fukugita1998} (hereafter named fixed case). Then, we will consider it  a free parameter to be constrained (free case). We perform a Monte Carlo Markov Chain (MCMC) analysis using the \textit{emcee} package \cite{Foreman-Mackey2013} to constrain the free parameters in our analysis: $m_{\gamma}$ and $\mathrm{DM}_{\mathrm{host,0}}$ (for fixed case), and $m_{\gamma}$, $f_{\mathrm{IGM,0}}$ and $\mathrm{DM}_{\mathrm{host,0}}$ (for free case). Since we are interested in a model-independent approach and to be consistent with our choice of $\mathrm{M}_{\mathrm{B}}$, we adopted the value for the Hubble constant in Eq. \ref{eq:constant} from SH0ES collaboration, $H_{0} = 74.03 \pm 1.4$ km/s/Mpc \cite{Riess2019}. We also assume the baryon density parameter, $\Omega_{b}h^{2} =  0.02235 \pm 0.00037$, from Big Bang Nucleosynthesis (BBN) analysis reported by \cite{Cooke2018}.

\begin{figure*}
\begin{center}
\includegraphics[width=0.46\textwidth]{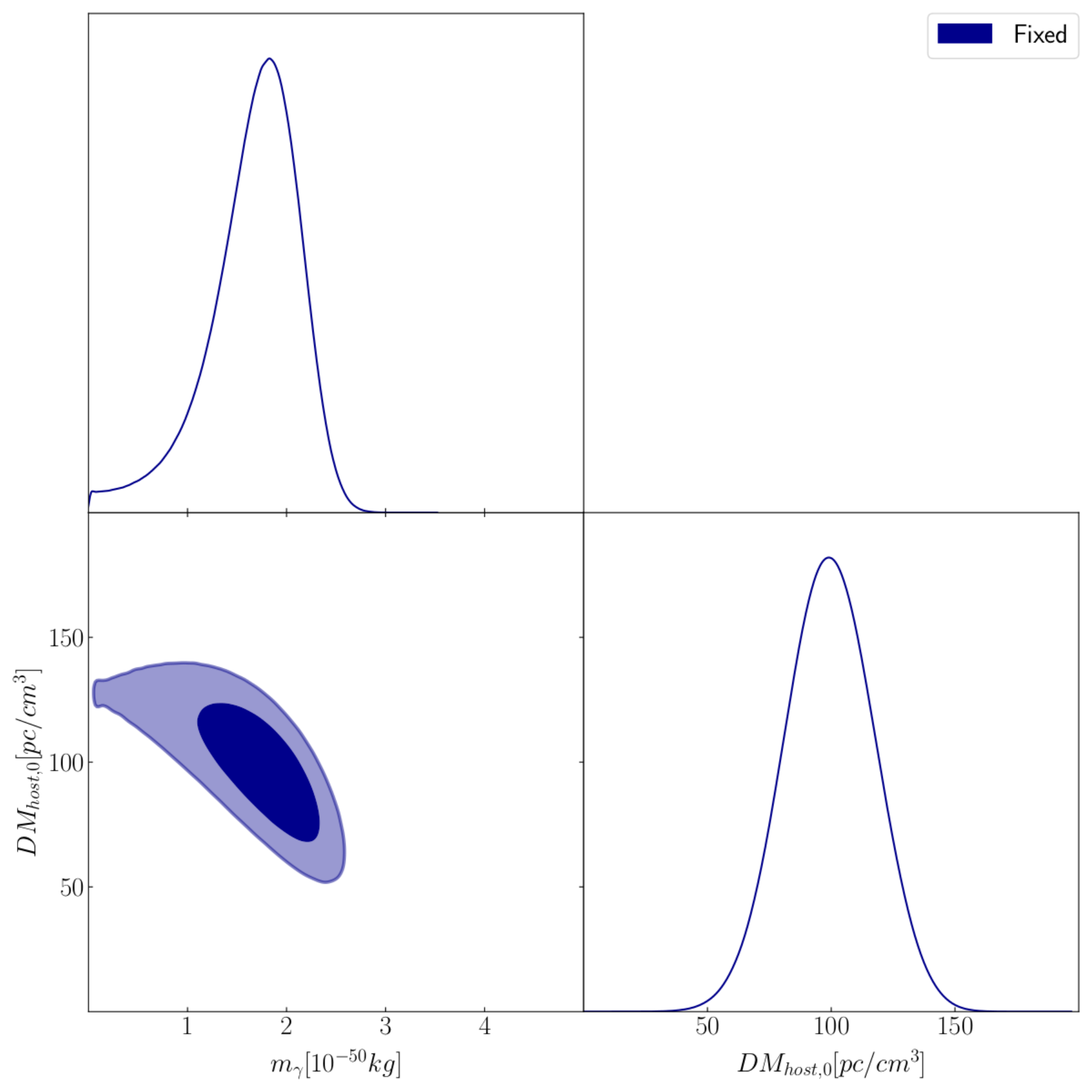}
\includegraphics[width=0.46\textwidth]{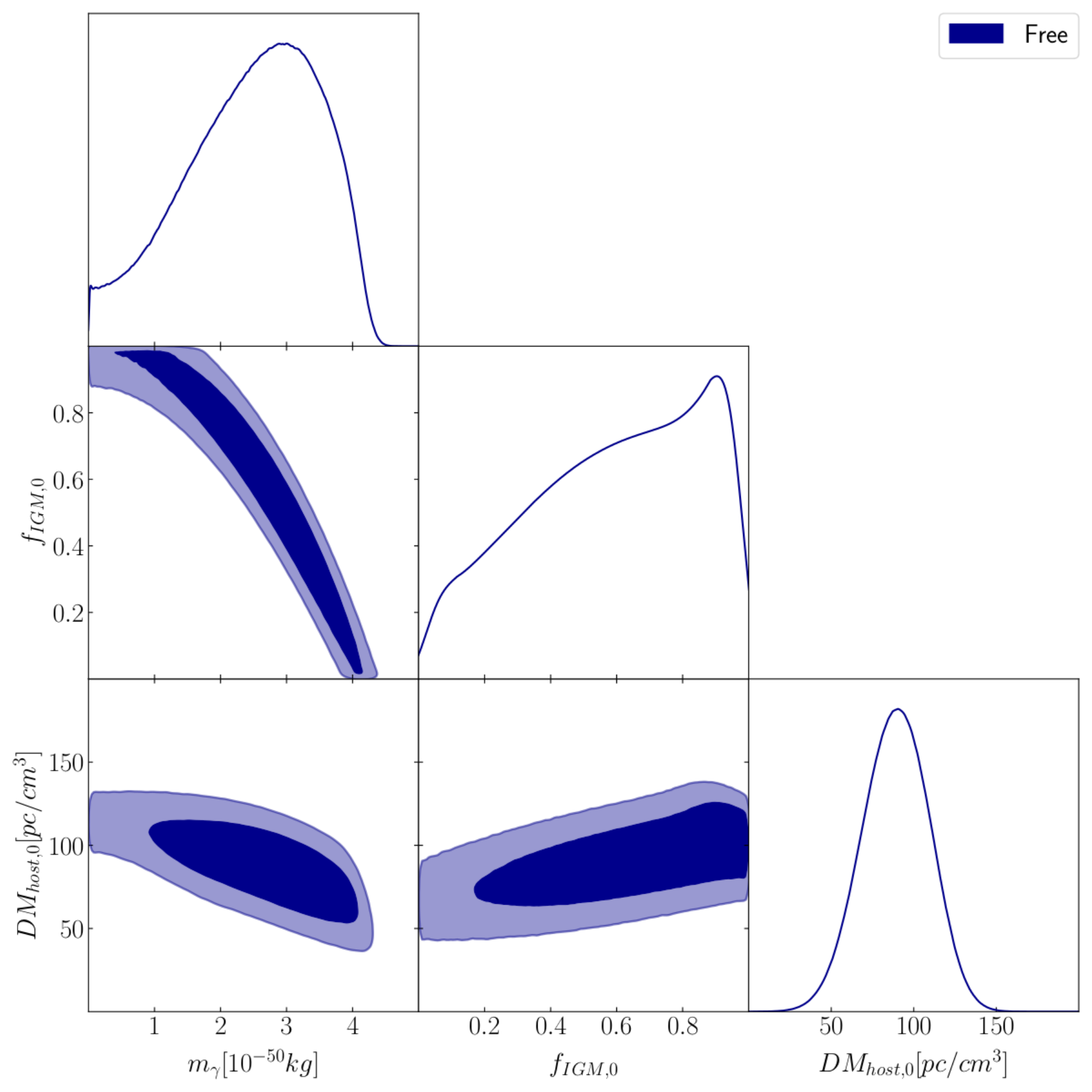}
\caption{Constraints on the photon rest mass $m_{\gamma}$, the baryon fraction $f_{\mathrm{IGM}}$ and the host galaxy contribution $\mathrm{DM}_{\mathrm{host},0}$ for the cases: Fixed $f_{\mathrm{IGM}}$ ({\it{left}}) and Free $f_{\mathrm{IGM}}$ ({\it{right}}).} 
\label{fig:results_obs}
\end{center} 
\end{figure*}

\section{Results}\label{sec:results}
\indent



Figure \ref{fig:results_obs} (right panel) shows the parametric space from the posterior probability density function and $1 - 2 \sigma$ contours for combinations of the parameters $m_{\gamma}$ and $\mathrm{DM}_{\mathrm{host},0}$ for $f_{\mathrm{IGM,0}}$ fixed. We obtain $m_{\gamma} = (18.2_{-5.9}^{+2.7}) \times 10^{-51}$ kg for the photon mass and the estimate for the host galaxy contribution is $\mathrm{DM}_{\mathrm{host},0} = 100 \pm 18$ pc/cm$^{3}$, both at $1\sigma$ level. Note that our estimative for the $m_{\gamma}$ agrees with the previous studies in the literature (see \cite{Wu2016,Bonetti2016,Bonetti2017,Shao2017,Xing2019,Wei2020,Wang2021,Chang2023,Lin2023,Wang2023,Ran2024}). However, our result is only consistent with $m_{\gamma} = 0$ at $\sim 3\sigma$. We also performed some tests for other values of $f_{\mathrm{IGM,0}}$ reported in the literature, and our results showed a tight relation between $f_{\mathrm{IGM,0}}$ and $m_{\gamma}$. Therefore, in what follows, we consider $f_{\mathrm{IGM},0}$ as a free parameter to study its impact on the  $m_{\gamma}$ constraints.


The left panel of Figure \ref{fig:results_obs} shows the posterior probability density function and $1 - 2 \sigma$ contours for combinations of the parameters $m_{\gamma}$, $f_{\mathrm{IGM},0}$ and $\mathrm{DM}_{\mathrm{host},0}$. From this analysis, we estimate the photon mass at $m_{\gamma} = (29.4_{-15.5}^{+5.8}) \times 10^{-51}$ kg and $f_{\mathrm{IGM,0}} = 0.902_{-0.631}^{+0.034}$ for the baryon fraction in IGM, both at $1\sigma$ level. We also estimate the host galaxy
contribution at $\mathrm{DM}_{\mathrm{host},0} = 90^{+19}_{-22}$ pc/cm$^{3}$ ($1\sigma$). This result differs from the  case in which $f_{\mathrm{IGM,0}}$ is fixed. The result obtained for photon mass is consistent with $m_{\gamma} = 0$ at $2\sigma$ and agrees with the other works in the literature, including Ref. \cite{Ran2024}, where the authors also use a cosmological independent method using a sample with 32 well-localized FRBs.

As commented earlier, any change in $m_{\gamma}$ could lead to a change in the other two parameters. Comparing the best-fit results for $m_{\gamma}$ in both cases of $f_{\mathrm{IGM,0}}$ (with a discrepancy value between them at $11.2 \times 10^{-31}$ kg), we note how much $f_{\mathrm{IGM,0}}$ is impacting the constraints on $m_{\gamma}$. From the contours in the left panel of Fig. \ref{fig:results_obs}, we see that the parameters $m_{\gamma}$ and $f_{\mathrm{IGM}}$ are strongly anticorrelated ($\mathrm{DM}_{\mathrm{host},0}$ is also correlated with the others parameters, but is not significant enough). We believe that such correlation between both parameters $m_{\gamma}$ and $f_{\mathrm{IGM}}$ comes from the equation of $\mathrm{DM}_{\mathrm{ext}}^{\mathrm{th}}$ (Eq. \ref{eq:DMext_th}), which is written in terms of components $\mathrm{DM}_{\mathrm{host}}$ ($\propto \mathrm{DM}_{\mathrm{host}} $), $\mathrm{DM}_{\mathrm{IGM}}$ ($\propto f_{\mathrm{IGM}}$) and $\mathrm{DM}_{\mathrm{\gamma}}$ ($\propto m_{\gamma}^{2}$).  



\section{Conclusions}\label{sec:conclusions}
\indent 

Several studies have reported constraints on the photon mass ($m_{\gamma}$) derived from FRBs data. However, almost all of these studies rely on the $\Lambda$CDM model, which introduces a circularity problem in the analysis, as the constancy of the speed of light is already assumed in standard cosmology.

In the present paper, we presented a cosmological model-independent approach to constrain the photon rest mass based on the dispersion measure of FRBs, combining 68 well-localized FRBs and the Pantheon SNe compilation. For the Fixed case (where $f_{\mathrm{IGM,0}}=0.83$), we found $m_{\gamma} = (18.2_{-5.9}^{+2.7}) \times 10^{-51}$ kg and $\mathrm{DM}_{\mathrm{host},0} = 100 \pm 18$ pc/cm$^{3}$ ($1\sigma$ level) whereas for the Free case obtained $m_{\gamma} = (29.4_{-15.5}^{+5.8}) \times 10^{-51}$ kg, $f_{\mathrm{IGM,0}} = 0.902_{-0.631}^{+0.034}$ and $\mathrm{DM}_{\mathrm{host},0} = 90^{+19}_{-22}$ pc/cm$^{3}$ at $1\sigma$ level. These results reflect the tight correlation between the parameters $m_{\gamma}$ and $f_{\mathrm{IGM}}$. Our estimated photon mass is larger than that presented by  \cite{Ran2024}, which found $m_{\gamma} \leq 3.5 \times 10^{-51}$ kg also using a cosmological model-independent approach. In our methodology, however, we only constrain three parameters compared to the six free parameters used in the other study. Reference \cite{Wang2023} also estimates $m_{\gamma}$ and $f_{\mathrm{IGM}}$ under the assumption of a fiducial $\Lambda$CDM cosmology with seven free parameters, but their results do not show a correlation between these parameters. The differences in results could be attributed to the significantly different sample sizes used in these studies, as the works mentioned use samples of 32 and 23 well-localized events. Additionally, the number of free parameters can affect the outcomes; increasing the degrees of freedom often leads to a better fit for the parameters.

Finally, these results emphasize the significant role of $f_{\mathrm{IGM}}$ in determining the photon mass. They also highlight the necessity of searching for a larger sample of FRBs and improving our understanding of their physical properties.


\section*{Acknowledgements}

TL thanks the financial support from the Conselho Nacional de Desenvolvimento Cient\'{\i}fico e Tecnol\'ogico (CNPq). RSG thanks financial support from the Funda\c{c}\~ao de Amparo \`a Pesquisa do Estado do Rio de Janeiro (FAPERJ) grant SEI-260003/005977/2024 - APQ1. JSA is supported by Conselho Nacional de Desenvolvimento Cient\'{\i}fico e Tecnol\'ogico (CNPq No. 307683/2022-2) and Funda\c{c}\~ao de Amparo \`a Pesquisa do Estado do Rio de Janeiro (FAPERJ) grant 259610 (2021). This work was developed thanks to the High-Performance Computing Center at the National Observatory (CPDON).








\newpage
\begin{longtable}[c]{l c c c c l}
\caption{Properties of FRBs}\\
\hline
\hline
\hline
Name & $z$ & $\mathrm{DM}_{\mathrm{ISM}}$ & $\mathrm{DM}_{\mathrm{obs}}$ & $\sigma_{\mathrm{obs}}$ & Refs. \\
& & [pc/cm$^{3}$] & [pc/cm$^{3}$] & [pc/cm$^{3}$] & \\
\hline
\endfirsthead

\hline
\hline
\hline
Name & $z$ & $\mathrm{DM}_{\mathrm{ISM}}$ & $\mathrm{DM}_{\mathrm{obs}}$ & $\sigma_{\mathrm{obs}}$ & Refs. \\
& & [pc/cm$^{3}$] & [pc/cm$^{3}$] & [pc/cm$^{3}$] & \\
\hline
\endhead

\hline
\endfoot

\hline
FRB 20121102A	&	0.19273	&	188.0	&	557.0	&	2.0	&	\cite{FRB121102}	\\
FRB 20180301A	&	0.3305	&	152.0	&	536.0	&	8.0	&	\cite{FRB20191228}	\\
FRB 20180814	&	0.068	&	87.75	&	189.4	&	0.4	&	\cite{FRB180814}	\\
FRB 20180916B	&	0.0337 	&	200.0	&	348.80	&	0.2	&	\cite{FRB180916}	\\
FRB 20180924B	&	0.3214	&	40.5	&	361.42	&	0.06	&	\cite{FRB180924}	\\
FRB 20181112A	&	0.4755	&	102.0	&	589.27	&	0.03	&	\cite{FRB181112}	\\
FRB 20181220A	&	0.2746	&	122.81	&	208.66	&	1.62	&	\cite{FRB181220A}	\\
FRB 20181223C	&	0.03024	&	19.9	&	112.45	&	0.01	&	\cite{FRB181220A}	\\
FRB 20190102C	&	0.2913	&	57.3	&	363.6	&	0.3	&	\cite{FRB190102}	\\
FRB 20190110C	&	0.12244	&	37	&	221.92	&	0.01	&	\cite{FRB190110C}	\\
FRB 20190303A	&	0.064	&	29.39	&	222.4	&	0.7	&	\cite{FRB180814}	\\
FRB 20190418A	&	0.07132	&	70.2	&	182.78	&	1.62	&	\cite{FRB181220A}	\\
FRB 20190425A	&	0.03122	&	49.25	&	127.78	&	1.62	&	\cite{FRB181220A}	\\
FRB 20190523A	&	0.66	&	37.0	&	760.8	&	0.6	&	\cite{FRB190523_1,FRB190523_2}	\\
FRB 20190608B	&	0.1178	&	37.2	&	338.7	&	0.5	&	\cite{FRB190608}	\\
FRB 20190611B	&	0.378	&	57.83	&	321.4	&	0.2	&	\cite{FRB190523_2}	\\
FRB 20190711A	&	0.522	&	56.4	&	593.1	&	0.4	&	\cite{FRB190523_2}	\\
FRB 20190714A	&	0.2365	&	38.0	&	504.13	&	2.0	&	\cite{FRB190523_2}	\\
FRB 20191001A	&	0.234	&	44.7	&	506.92	&	0.04	&	\cite{FRB190523_2}	\\
FRB 20191106C	&	0.10775	&	25	&	333.40	&	0.2	&	\cite{FRB190110C}	\\
FRB 20191228A	&	0.2432	&	33.0	&	297.5	&	0.05	&	\cite{FRB20191228}	\\
FRB 20200223B	&	0.06024	&	46	&	202.268	&	0.007	&	\cite{FRB190110C}	\\
FRB 20200430A	&	0.16	&	27.0	&	380.25	&	0.5	&	\cite{FRB190523_2}	\\
FRB 20200906A	&	0.3688	&	36	&	577.8	&	0.02	&	\cite{FRB20191228}	\\
FRB 20201123A	&	0.0507	&	251.93	&	433.55	&	0.0036	&	\cite{FRB201123A}	\\
FRB 20201124A	&	0.098	&	123.2	&	413.52	&	0.5	&	\cite{FRB201124}	\\
FRB 20210117A	&	0.2145	&	34.4	&	730.0	&	1.0	&	\cite{FRB210117A}	\\
FRB 20210320	&	0.2797	&	42.2	&	384.8	&	0.3	&	\cite{FRB210320}	\\
FRB 20210410D	&	0.1415	&	56.2	&	578.78	&	2.0	&	\cite{FRB210410D}	\\
FRB 20210603A 	&	0.1772	&	40.0	&	500.147	&	0.004	&	\cite{FRB210603A}	\\
FRB 20210807D	&	0.12927	&	121.2	&	251.9	&	0.2	&	\cite{FRB210807D}	\\
FRB 20211127I	&	0.0469	&	42.5	&	234.83	&	0.08	&	\cite{FRB210807D}	\\
FRB 20211203C 	&	0.3439	&	63.4	&	636.2	&	0.4	&	\cite{FRB211203C}	\\
FRB 20211212A	&	0.0715	&	27.1	&	206.0	&	5.0	&	\cite{FRB210807D}	\\
FRB 20220105A	&	0.2785	&	22.0	&	583	&	1.0	&	\cite{FRB211203C}	\\
FRB 20220204A	&	0.4	&	50.7	&	612.2	&	0.05	&	\cite{FRB220204A}	\\
FRB 20220207C	&	0.043040	&	79.3 	&	262.38	&	0.01	&	\cite{FRB220207C}	\\
FRB 20220208A	&	0.351	&	101.6	&	437	&	0.6	&	\cite{FRB220204A}	\\
FRB 20220307B	&	0.248123	&	135.7 	&	499.27 	&	0.06	&	\cite{FRB220207C}	\\
FRB 20220310F	&	0.477958	&	45.4 	&	462.24	&	0.005	&	\cite{FRB220207C}	\\
FRB 20220330D 	&	0.3714	&	38.6	&	468.1	&	0.85	&	\cite{FRB220204A}	\\
FRB 20220418A	&	0.622000	&	37.6 	&	623.25	&	0.01	&	\cite{FRB220207C}	\\
FRB 20220501C	&	0.381	&	31	&	449.5	&	0.2	&	\cite{FRB210807D}	\\
FRB 20220506D	&	0.30039 	&	89.1	&	396.97	&	0.02	&	\cite{FRB220207C}	\\
FRB 20220509G	&	0.089400	&	55.2	&	269.53	&	0.02	&	\cite{FRB220207C}	\\
FRB 20220610A	&	1.016	&	31.0	&	1458.15	&	0.2	&	\cite{FRB20220610A}	\\
FRB 20220717A	&	0.36295	&	118	&	637.34	&	3.52	&	\cite{FRB220717A}	\\
FRB 20220725A	&	0.1926	&	31	&	290.4	&	0.3	&	\cite{FRB210807D}	\\
FRB 20220726A	&	0.361	&	89.5	&	686.55	&	0.01	&	\cite{FRB220204A}	\\
FRB 20220825A	&	0.241397	&	79.7	&	651.24	&	0.06	&	\cite{FRB220207C}	\\
FRB 20220831A 	&	0.262	&	1019.50	&	1146.25	&	0.2	&	\cite{FRB220204A}	\\
FRB 20220912A	&	0.0771	&	125.00	&	219.46	&	0.042	&	\cite{FRB220912A}	\\
FRB 20220914A	&	0.113900 	&	55.2	&	631.28	&	0.04	&	\cite{FRB220207C}	\\
FRB 20220918A	&	0.491	&	41	&	656.8	&	0.8	&	\cite{FRB210807D}	\\
FRB 20220920A	&	0.158239	&	40.3	&	314.99 	&	0.01	&	\cite{FRB220207C}	\\
FRB 20221012A	&	0.284669	&	54.4	&	441.08	&	0.7	&	\cite{FRB220207C}	\\
FRB 20221106A	&	0.2044	&	35	&	343.8	&	0.8	&	\cite{FRB210807D}	\\
FRB 20221219A 	&	0.554	&	44.4	&	706.7	&	0.6	&	\cite{FRB221219A}	\\
FRB 20230526A	&	0.1570	&	50	&	361.4	&	0.2	&	\cite{FRB210807D}	\\
FRB 20230708A	&	0.105	&	50	&	411.51	&	0.05	&	\cite{FRB210807D}	\\
FRB 20230718A	&	0.035	&	396	&	477.0	&	0.5	&	\cite{FRB210807D}	\\
FRB 20230814A	&	0.5535	&	104.9	&	696.35	&	0.5	&	\cite{FRB220204A}	\\
FRB 20230902A	&	0.3619	&	34	&	440.1	&	0.1	&	\cite{FRB210807D}	\\
FRB 20231226A	&	0.1569	&	145	&	329.9	&	0.1	&	\cite{FRB210807D}	\\
FRB 20240114A	&	0.13	&	49.7	&	527.65	&	0.01	&	\cite{FRB240114A}	\\
FRB 20240201A	&	0.042729	&	38	&	374.5	&	0.2	&	\cite{FRB210807D}	\\
FRB 20240210A	&	0.023686	&	31	&	283.73	&	0.05	&	\cite{FRB210807D}	\\
FRB 20240310A	&	0.1270	&	36	&	601.8	&	0.2	&	\cite{FRB210807D}	\\
\label{tab:data}
\end{longtable}

\end{document}